\documentclass[10pt,preprint]{aastex}
\usepackage{graphicx,apjfonts,emulateapj5,onecolfloat5}
\usepackage{subfigure}
\usepackage{ulem}
\usepackage{epsf}
\usepackage{dcolumn}   
\usepackage{bm}        
\usepackage{color}

\usepackage{threeparttable}
\usepackage{tabularx}

\shorttitle{Hubble constant from Hubble parameter data}
\shortauthors{Chen, Y. et al.}

\begin{document}
\twocolumn[


\title{Determining the Hubble constant from Hubble parameter measurements}


\author{Yun Chen\altaffilmark{1},
        Suresh Kumar\altaffilmark{2},
        Bharat Ratra\altaffilmark{3}
        }


\begin{abstract}
We use 28 Hubble parameter, $H(z)$, measurements at intermediate redshifts
$0.07 \leq z \leq 2.3$ to determine the present-day Hubble constant $H_0$
in four cosmological models. We measure $H_0 = 68.3^{ +2.7}_{ -2.6  },
68.4^{ +2.9   }_{ -3.3   }, 65.0^{ +6.5    }_{ -6.6    }$ and
$ 67.9^{ +2.4}_{-2.4}$ km s${}^{-1}$ Mpc${}^{-1}$ (1$\sigma$ errors) in the
$\Lambda$CDM (spatially flat and non-flat), $\omega$CDM and $\phi$CDM models, respectively.
These measured $H_0$ values are more consistent with the lower values
determined from recent cosmic microwave background and baryon acoustic
oscillation data, as well as with that found from a median statistics analysis
of Huchra's compilation of $H_0$ measurements,but include the higher local measurements of $H_0$ within the 2$\sigma$ confidence limits.

\end{abstract}


]

\altaffiltext{1}{Key Laboratory for Computational Astrophysics, National Astronomical Observatories, Chinese Academy of Sciences, Beijing, 100012, China; chenyun@bao.ac.cn}
\altaffiltext{2}{Department of Mathematics, BITS Pilani, Pilani Campus, Rajasthan-333031, India; suresh.kumar@pilani.bits-pilani.ac.in}
 \altaffiltext{3}{Department of Physics, Kansas State University, 116 Cardwell Hall, Manhattan, KS
66506, USA; ratra@phys.ksu.edu}


\section{Introduction}

The current value of the cosmological expansion rate, the Hubble constant
$H_0$, is an important cosmological datum. Although one of the most measured
cosmological parameters, it was more than seven decades after Hubble's
first measurement before a consensus value for $H_0$ started to emerge.
In 2001 Freedman et al. (2001) provided $H_0 = 72 \pm 8$ km
s${}^{-1}$ Mpc${}^{-1}$ (1$\sigma$ error including systematics) as a
reasonable summary of the Hubble Space Telescope Key Project $H_0$ value.
In the same year Gott et al. (2001) applied median statistics\footnote{
For applications and discussions of median statistics see Podariu et al.
(2001), Chen \& Ratra
(2003), Mamajek \& Hillenbrand (2008), Croft \& Dailey (2015),
Andreon \& Hurn (2012), Farooq et al. (2013a), Crandall \& Ratra (2014, 2015),
Ding et al. (2015), Crandall et al. (2015), and Zheng et al. (2016). Median
statistics does not make use of the error bars of the individual mesaurements.}
to 331 $H_0$ estimates tabulated by Huchra\footnote
{https://www.cfa.harvard.edu/$\sim$dfabricant/huchra/}
and determined $H_0 = 67 \pm 3.5$
km s${}^{-1}$ Mpc${}^{-1}$. During the following decade median statistics was
applied to larger compilations of $H_0$ measurements from Huchra, in 2003
to 461 measurements by Chen et al. (2003) who found $H_0 = 68 \pm 3.5$
km s${}^{-1}$ Mpc${}^{-1}$, and in 2011 to 553 measurements by Chen \&
Ratra (2011) who found $H_0 = 68 \pm 2.8$ km s${}^{-1}$ Mpc${}^{-1}$.

Many more recent $H_0$ determinations are consistent with these results. For
instance, the final Wilkinson Microwave Anisotropy Probe (WMAP) measurement is
$H_0 = 70.0 \pm 2.2$ km s${}^{-1}$ Mpc${}^{-1}$ (Hinshaw et al. 2013),
while the Atacama Cosmology Telescope and the WMAP 7-year cosmic
microwave background (CMB) anisotropy data give $H_0 = 70.0 \pm 2.4$
km s${}^{-1}$ Mpc${}^{-1}$ (Sievers et al. 2013), and baryon acoustic
oscillations (BAO), type Ia supernovae, and CMB data result in
$H_0 = 67.3 \pm 1.1$ km s${}^{-1}$ Mpc${}^{-1}$ (Aubourg et al. 2015;
also see Ross et al. 2015; L 'Huillier \& Shafieloo 2016; Bernal et al. 2016;
Lukovi{\'c} et al. 2016), with the
Planck 2015 CMB data value being $H_0 = 67.8 \pm 0.9$ km s${}^{-1}$
Mpc${}^{-1}$ (Ade et al. 2015; but see Addison et al. 2016).

While the consistency of these results are encouraging, some recent local
estimates of $H_0$ are larger. Riess et al. (2011) find $H_0 = 73.8 \pm
2.4$ km s${}^{-1}$ Mpc${}^{-1}$ (but see Efstathiou 2014 who argues that
$H_0 = 72.5 \pm 2.5$ km s${}^{-1}$ Mpc${}^{-1}$ is a better representation),
Freedman et al. (2012) find $H_0 = 74.3 \pm 2.1$ km s${}^{-1}$ Mpc${}^{-1}$
while Riess et al. (2016) give $H_0 = 73.24 \pm 1.74$ km s${}^{-1}$
Mpc${}^{-1}$.

It is important to understand the reasons for this difference. For instance,
the value and uncertainty of $H_0$ affects observational constraints on other
cosmological parameters (see, e.g., Samushia et al. 2007; Chen et al. 2016);
given current cosmological data, the standard model of particle physics with
three light neutrino species is
more compatible with the lower $H_0$ value and difficult to reconcile with
the higher value (see, e.g., Calabrese et al.
2012); and the difference between the local and global $H_0$ values might be
an indication that the $\Lambda$CDM model needs to be extended (see, e.g.,
Di Valentino et al. 2016).

Here we use Hubble parameter, $H(z)$ (where $z$ is redshift), measurements to
determine the Hubble constant. $H(z)$ data have previously been used to
constrain other cosmological parameters (see, e.g., Samushia \& Ratra 2006;
Chen \& Ratra 2011b; Farooq et al. 2013b, 2015; Farooq \& Ratra 2013a;
Capozziello et al. 2014; Chen et al. 2015; Meng et al. 2015;
Guo \& Zhang 2016; Sol\`a et al. 2016; Alam et al. 2016; Mukherjee 2016),
including measuring the redshift of the
cosmological deceleration-acceleration transition between the earlier
nonrelativistic matter dominated epoch and the current dark energy dominated
epoch (see, e.g., Farooq \& Ratra 2013b;
Moresco et al. 2016). See Verde et al. (2014) for an early attempt at
measuring $H_0$ from $H(z)$ data. Here we use more data (28 vs. 15
measurements) to higher redshift (2.30 vs. 1.04) than Verde et al. (2014)
used and so find tighter constraints on $H_0$.

We find that our $H(z)$ $H_0$ values are more
consistent with the lower values determined using median statistics or
from CMB anisotropy or BAO measurements and with the predictions of the
standard model of particle physics with only three light neutrino species
and no ``dark radiation''. Systematic errors affecting $H(z)$ measurements
are largely different from those affecting CMB and BAO measurements. In
addition, median statistics does not make use of the error bars of the
individual measurements. It is significant that all four techniques
result in very similar values of $H_0$.

To determine $H_0$ we analyze the $H(z)$ data tabulated in Farooq \& Ratra
(2013b) and reproduced in Table 1 here\footnote{The error bars of these $H(z)$
measurements include systematic errors. In the analyses here we ignore the
correlations between the 3 Blake et al. (2012) points; these only very
slightly affect the results (Farooq et al. 2016).}, using two different
dark energy models, $\Lambda$CDM (Peebles 1984) and $\phi$CDM
(Peebles \& Ratra 1988; Ratra \& Peebles 1988), as well as an incomplete,
but popular, parameterization of dark energy, $\omega$CDM.
In all cases we measure $H_0$ from the one-dimensional likelihood determined by
marginalizing over all other parameters. (Limits on other parameters, such
as the current nonrelativistic matter density parameter, are quite reasonable.)

In the next section we summarize the models we use, as well as the $\omega$CDM
parametrization. In Sec.\ 3 we present our $H_0$ determinations. We conclude
in the final section.

\begin{table}
\begin{center}
\caption{Hubble parameter versus redshift data.}
\begin{tabular}{cccc}
\hline\hline
$z$ & $H(z)$ &$\sigma_{H}$  &Reference\tnote{a}\\
    & (km s$^{-1}$ Mpc $^{-1}$) & (km s$^{-1}$ Mpc $^{-1}$)& \\
\tableline\\[-4pt]
0.070&	69&	19.6& 5\\
0.090&	69&	12&	1\\
0.120&	68.6&	26.2&	5\\
0.170&	83&	8&	1\\
0.179&	75&	4&	3\\
0.199&	75&	5&	3\\
0.200&	72.9&	29.6&	5\\
0.270&	77&	14&	1\\
0.280&	88.8&	36.6&	5\\
0.350&	76.3&	5.6&	7\\
0.352&	83&	14&	3\\
0.400&	95&	17&	1\\
0.440&	82.6&	7.8&	6\\
0.480&	97&	62&	2\\
0.593&	104&	13&	3\\
0.600&	87.9&	6.1&	6\\
0.680&	92&	8&	3\\
0.730&	97.3&	7.0&	6\\
0.781&	105&12&	3\\
0.875&	125&	17&	3\\
0.880&	90&40&	2\\
0.900&	117&	23&	1\\
1.037&	154&	20&	3\\
1.300&168&	17&	1\\
1.430&	177&	18&	1\\
1.530&	140&	14&	1\\
1.750&	202&	40&	1\\
2.300&	224&	8&	4\\
\hline\hline
\end{tabular}
\begin{tablenotes}
\item[a]{Reference numbers:
1. Simon et al. (2005), 2. Stern et al. (2010),
3. Moresco et al. (2012), 4. Busca et al. (2013),
5. Zhang et al. (2014), 6. Blake et al. (2012),
7. Chuang \& Wang (2013).}
\end{tablenotes}
\end{center}
\label{tab:Hz}
\end{table}

\section{$\Lambda$CDM, $\omega$CDM and $\phi$CDM} \label{sec:models}

The Hubble parameter of the spatially-flat $\Lambda$CDM model is
\begin{equation}\label{eq1}
H (z) = H_{0} \sqrt{\Omega_{m0}(1+z)^3+1-\Omega_{m0}},
\end{equation}
while in the general (non-flat) $\Lambda$CDM model it is
\begin{equation}\label{eq2}
H (z) = H_{0} \sqrt{\Omega_{m0}(1+z)^3 +
                    (1- \Omega_{m0} - \Omega_\Lambda) (1+z)^2 + \Omega_\Lambda},
\end{equation}
where $\Omega_{m0}$ is the current value of the nonrelativistic matter
density parameter and $\Omega_\Lambda$ is the cosmological constant
density parameter.

In the spatially-flat $\omega$CDM parametrization we have
\begin{equation}\label{eq3}
H(z) = H_{0}\sqrt{\Omega_{m0}(1+z)^3+(1-\Omega_{m0})(1+z)^{3(1+w_X)}},
\end{equation}
where $\omega_X$ is the constant, negative, equation of state parameter
relating the (dark energy) $X$-fluid pressure and energy density
through $p_X = \omega_X \rho_X$. The $\omega$CDM parametrization is
incomplete and does not consistently describe inhomogeneities. However,
$\phi$CDM, discussed next, is a consistent dynamical dark energy
model.

The Friedmann equation of the spatially-flat $\phi$CDM model is
\begin{equation}
\label{eq:phiCDMFriedmann} H^2(z) =
\frac{8\pi}{3m_p^2}(\rho_m + \rho_{\phi}),
\end{equation}
where $m_p$ is the Planck mass, $\rho_m$ is the nonrelativistic matter
energy density and the scalar field $\phi$ energy density is
\begin{equation}
\label{eq:rhophi} \rho_{\phi} = \frac{m_p^2}{32\pi} (\dot{\phi}^2 +
\kappa m_p^2 \phi^{-\alpha}).
\end{equation}
Here an overdot denotes a time derivative, $\kappa (m_p, \alpha)$ and
$\alpha$ are positive constants, and we have picked an inverse-power-law
scalar field potential energy density $V(\phi) = \kappa m_p^2
\phi^{-\alpha}/2$. The scalar field equation of motion is
\begin{equation}
\ddot\phi+3{{\dot a}\over a}\dot \phi+\frac{d V}{d\phi}=0
\label{eq:KGeq}
\end{equation}
where $a$ is the scale factor. These equations are numerically integrated
to provide $H(z)$ in the $\phi$CDM model (Peebles \& Ratra 1988; Samushia 2009;
Farooq 2013).

\section{Analysis and results}

We constrain cosmological
parameters by minimizing $\chi_{H}^2$,
\begin{equation}
\label{eq:chi2Hz} \chi_{H}^2 (\textbf{p}) =
\sum_{i=1}^{N}\frac{[H^{\rm th} (z_i; \textbf{p})-H^{\rm
obs}(z_i)]^2}{\sigma^2_{{\rm H},i}},
\end{equation}
for $N$ measured $H^{\rm obs} (z_i)$'s with variance $\sigma^2_{{\rm H},i}$
at redshift $z_i$ where $H^{\rm th}$ is the predicted value of $H(z)$
in the cosmological model. \textbf{p} represents the free parameters
of the cosmological model under consideration, $H_0$ and $\Omega_{m0}$ in all
four cases, with one additional parameter in three of the cases:
$\Omega_\Lambda$ in non-flat $\Lambda$CDM, $\omega_X$ in the spatially-flat
$\omega$CDM parameterization, and $\alpha$ in the spatially-flat
$\phi$CDM model. We use the compilation of 28 $H(z)$ data points from
Farooq \& Ratra (2013b) as reproduced here in Table 1 to constrain the
model parameters under consideration by using the Markov Chain Monte Carlo method coded in the
publicly available package \textbf{CosmoMC} (Lewis \& Bridle 2002).

Our results are summarized in Table 2 and Figs. 1--5.

\begin{table*}
\begin{center}
\caption{Mean values of free parameters of various models with 1$\sigma$ and 2$\sigma$ error bars.} \label{tab:results}
\begin{tabular}{ccccc}
\hline\hline
Parameter & $\Lambda$CDM & Non-flat $\Lambda$CDM & XCDM & $\phi$CDM \\
\hline \\[6pt]
$H_0$	& $68.3^{ +2.7 +5.2   }_{ -2.6 -5.1   }$ & $68.4^{ +2.9 +5.9   }_{ -3.3 -5.4   }$ & $65.0^{ +6.5 +9.4   }_{ -6.6 -9.3   }$ & $ 67.9^{ +2.4 +4.7}_{-2.4 -4.7}$ \\[6pt]
$\Omega_{m0}$ & $0.276^{ +0.032 +0.072   }_{ -0.039 -0.068   }$ & $0.267^{ +0.049 +0.010   }_{ -0.050 -0.102   }$ & $0.308^{ +0.048 +0.114   }_{ -0.076 -0.102   }$ & $ 0.275^{ +0.029 +0.063}_{ -0.035 -0.062}$ \\[6pt]
$\Omega_{\Lambda}$	&$--$&$ 0.708^{ +0.101 +0.219   }_{ -0.167 -0.208   }$ & $--$ & $--$ \\[6pt]
$w_X$	& $--$&$--$&$ -0.780^{ +0.196 +0.460   }_{ -0.292 -0.414   }$&$--$ \\[6pt]
$\alpha$	& $--$&$--$&$ --$& ${\rm no\ limits}$ \\[6pt]
 \hline\\
$\chi^2_{\rm min}$ & $17.0$ & $16.9$ & $17.0$ & $17.0$ \\[6pt]
\hline\hline
\end{tabular}
\end{center}
\end{table*}

\begin{figure}
\begin{center}
\includegraphics[width=0.8\hsize]{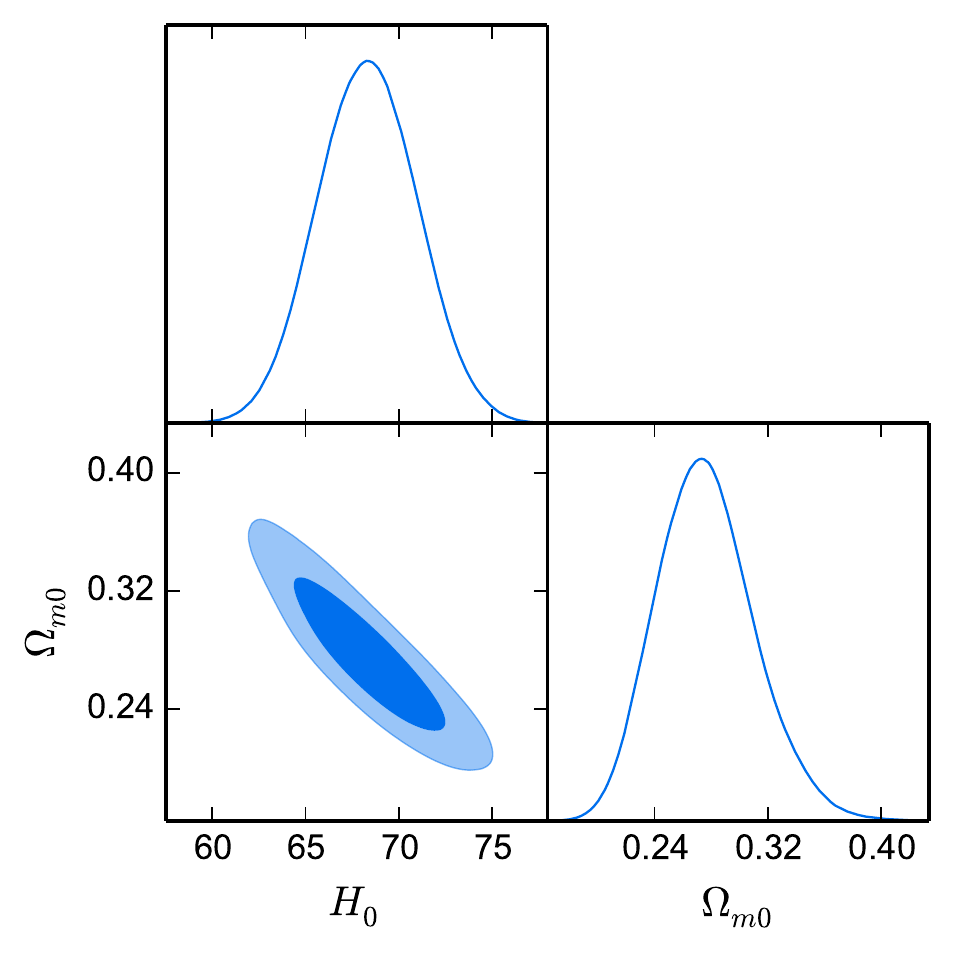}
\end{center}
\caption{1$\sigma$ and 2$\sigma$ confidence contours of spatially-flat $\Lambda$CDM model parameters. Marginalized probability distributions of the individual parameters are also displayed.}
\end{figure}

\begin{figure}
\begin{center}
\includegraphics[width=0.8\hsize]{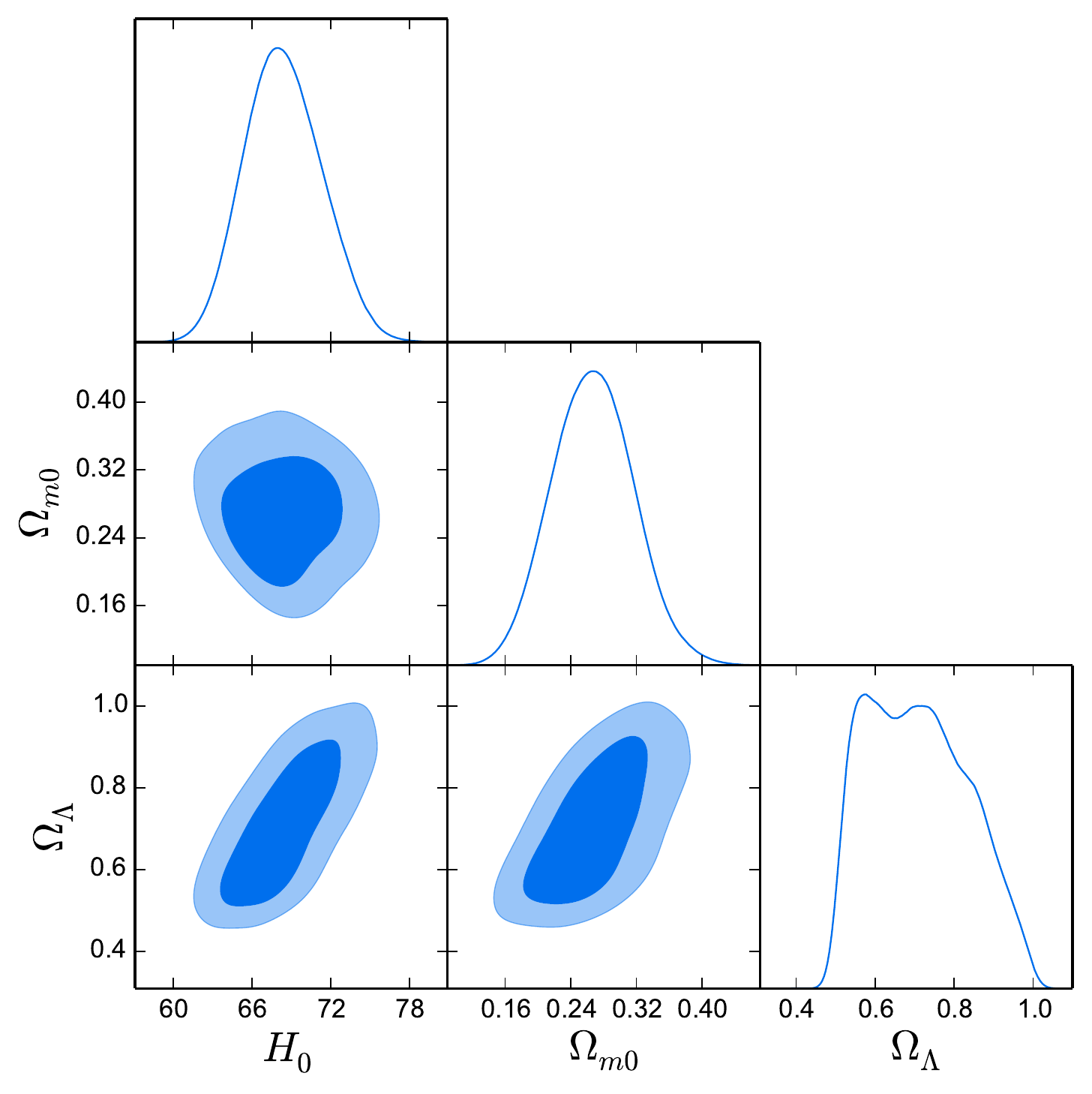}
\end{center}
\caption{1$\sigma$ and 2$\sigma$ confidence contours of non-flat $\Lambda$CDM model parameters. Marginalized probability distributions of the individual parameters are also displayed.}
\end{figure}

\begin{figure}
\begin{center}
\includegraphics[width=0.8\hsize]{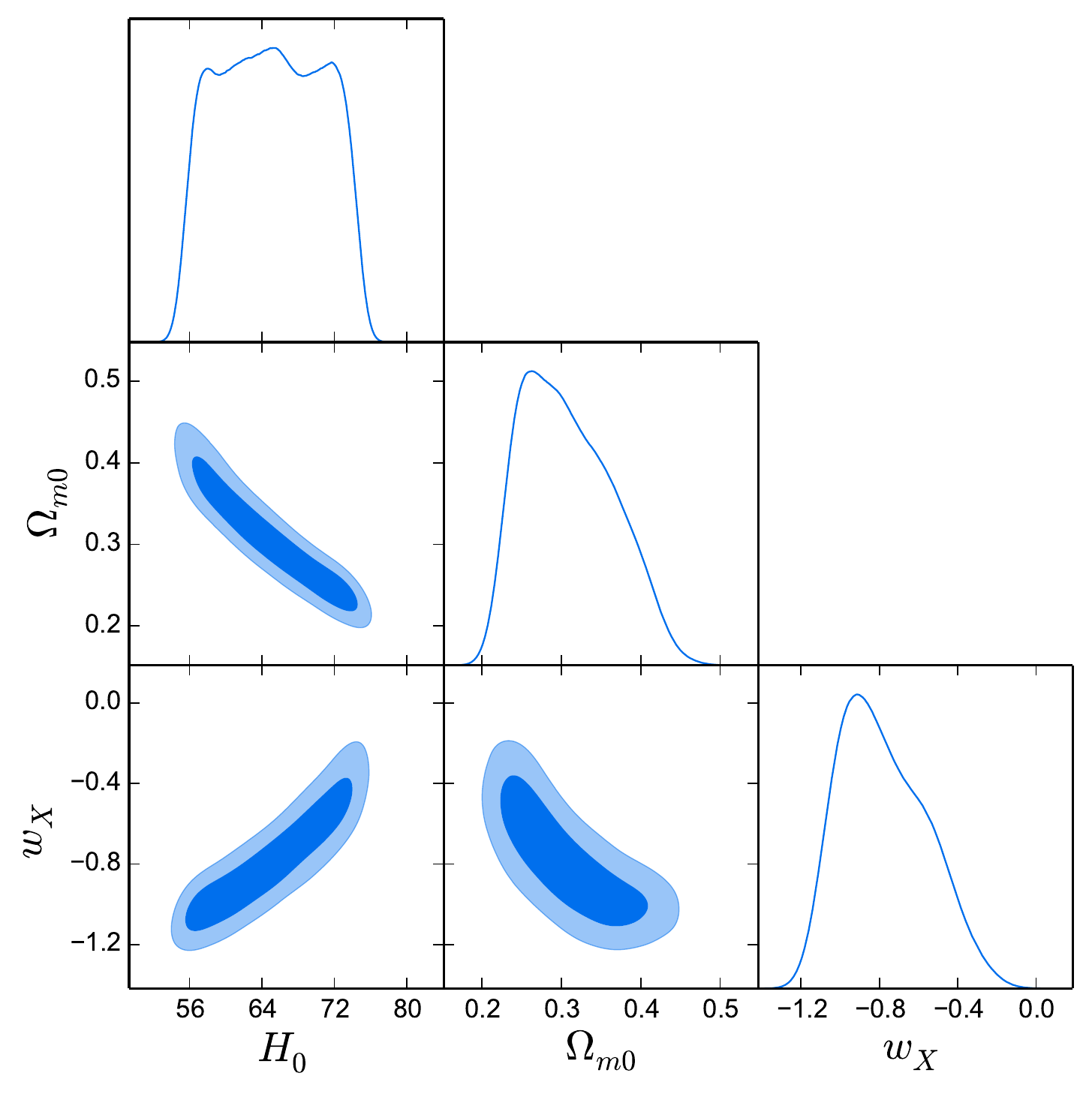}
\end{center}
\caption{1$\sigma$ and 2$\sigma$ confidence contours of the spatially-flat
$\omega$CDM parameterization parameters. Marginalized probability distributions of the individual parameters are also displayed.}
\end{figure}

\begin{figure}
\begin{center}
\includegraphics[width=0.8\hsize]{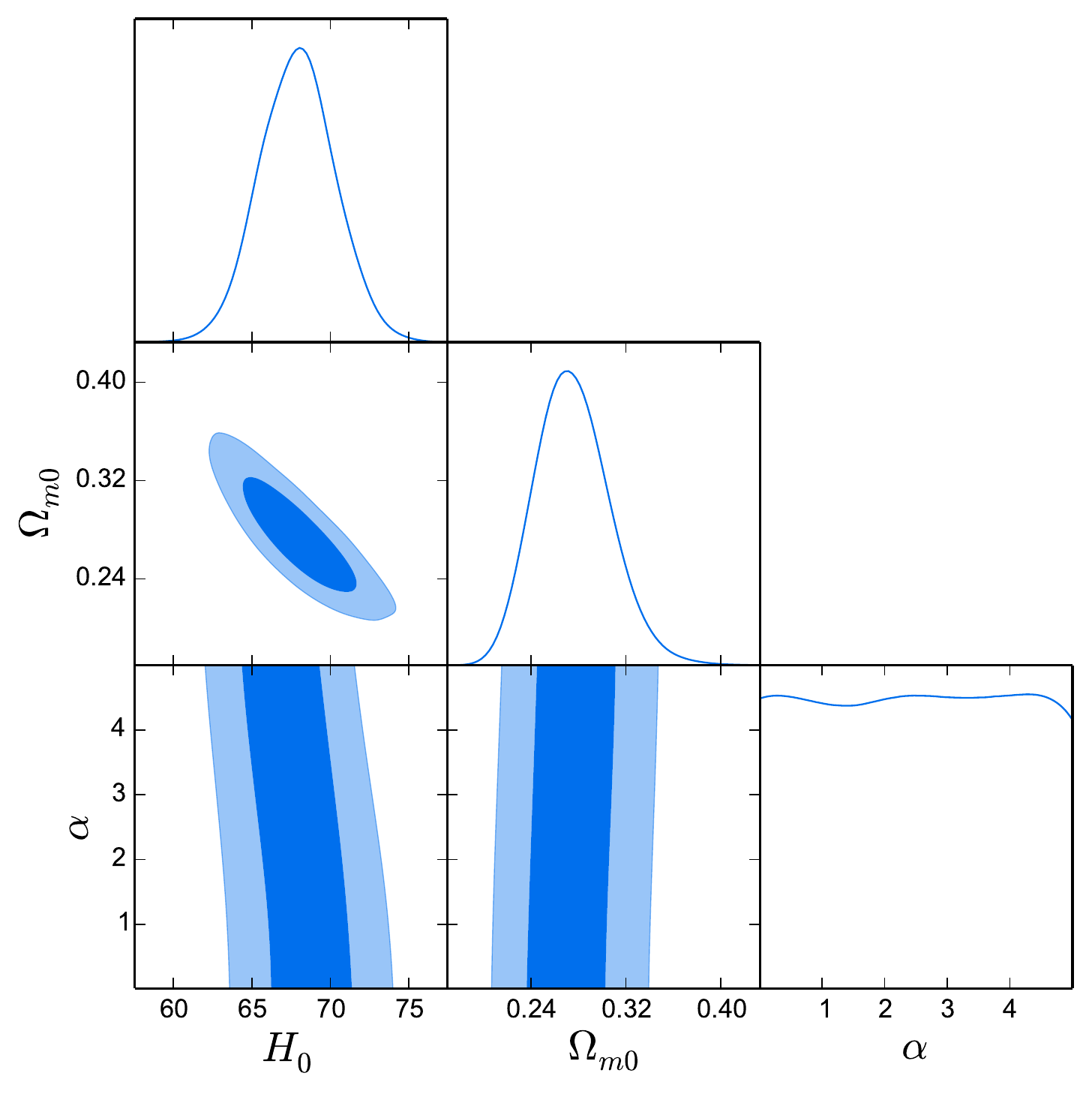}
\end{center}
\caption{1$\sigma$ and 2$\sigma$ confidence contours of the spatially-flat
$\phi$CDM model parameters. Marginalized probability distributions of the individual parameters are also displayed.}
\end{figure}

\begin{figure}
\begin{center}
\includegraphics[width=1\hsize]{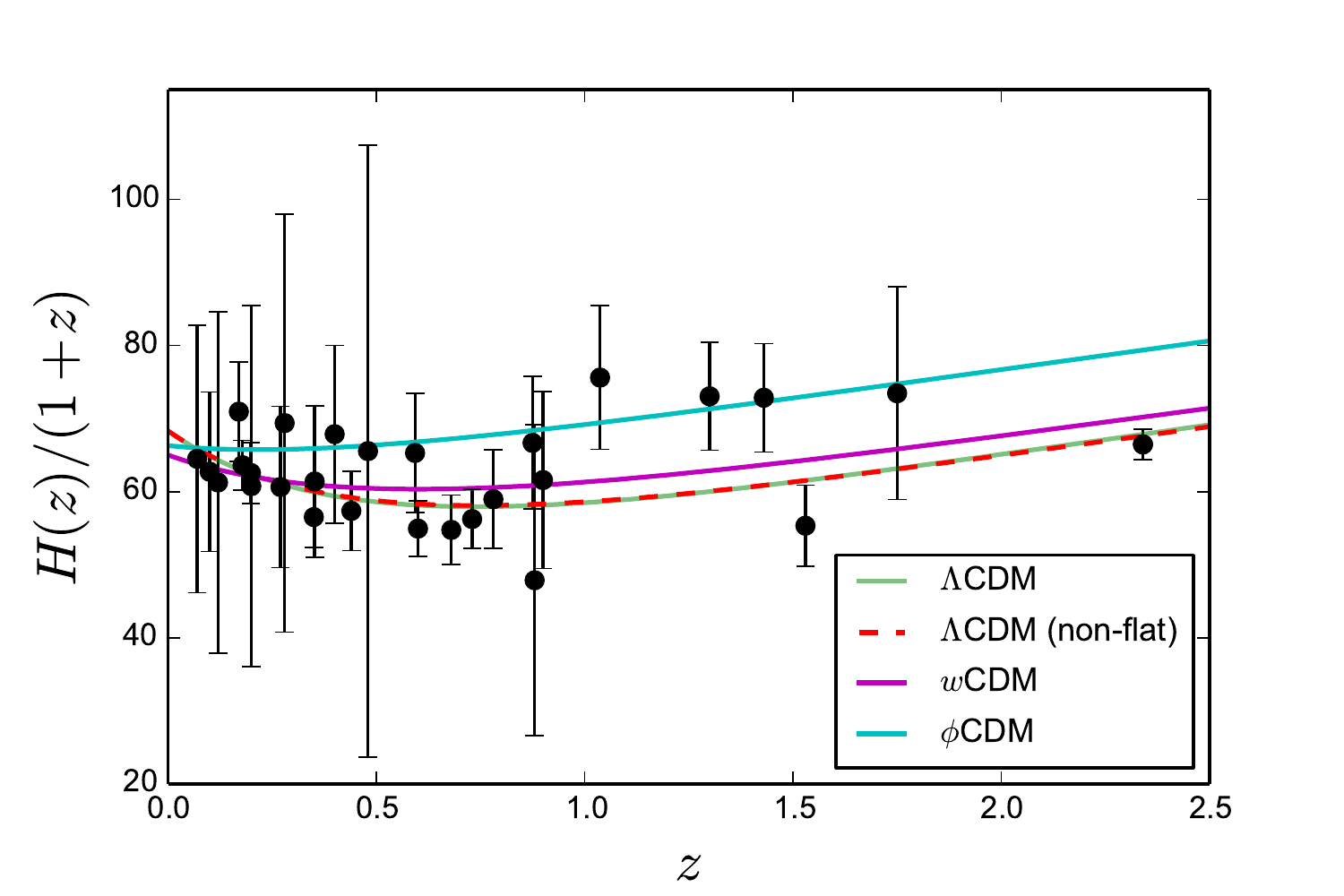}
\end{center}
\caption{Best-fit model curves and the 28 $H(z)$ data points.}
\end{figure}

The limits on cosmological parameters shown in Table 2 are derived from the
corresponding one-dimensional likelihood function that results from
marginalizing over all of the other parameters. The constraints
listed in Table 2 are roughly in line with those now under discussion.The
small reduced $\chi^2$'s which follow from the entries in the last line of
the Table are not unexpected given the results of Farooq et al. (2013a).

The $H_0$ values listed in Table 2 are in good accord with the lower recent
values determined by using median statistics on Huchra's compilation and from
CMB and BAO data as well as with what is expected in the standard model of
particle physics with only three light neutrino species and no additional
``dark radiation''.

There are two high-weight data subsets in our analysis: the cosmic chronometer
data from Moresco et al. (2012) and the Lyman-$\alpha$ data from Busca et al.
(2013). Since both of these results are based on relatively new approaches
to measuring $H(z)$, it is informative to see an analysis of $H_0$
when one and then the other of these data sets are omitted from the analysis.
When we drop the Moresco et al. (2012) data from the compilation, we find
$H_0=67.5^{+3.7+8.0}_{-3.7-8.0}$ km s${}^{-1}$ Mpc${}^{-1}$ while dropping
Busca et al. (2013) point we obtain $H_0=66.9^{+2.8+5.3}_{-2.8-5.5}$
km s${}^{-1}$ Mpc${}^{-1}$. Comparing these with the full-data
result $H_0=68.3^{ +2.7 +5.2   }_{ -2.6 -5.1   }$ km s${}^{-1}$ Mpc${}^{-1}$,
we observe a minor shift in the central values and larger error bars when
one or the other data subset is omitted from the compilation.

\section{Conclusions}

We have used the $H(z)$ data tabulated in Farooq \& Ratra (2013b) as reproduced here in Table 1 to measure
$H_0$. The $H_0$ values we find are more consistent with the lower values
determined from the recent CMB and BAO data, as well as with that found from a median statistics analysis
of Huchra's compilation of $H_0$ measurements.

\section*{Acknowledgments}

Y.C. was supported by the National Natural Science Foundation of China
(Nos. 11133003 and 11573031), the Strategic Priority Research Program ``The Emergence of Cosmological Structures'' of the Chinese Academy of Sciences
(No. XDB09000000), and the China Postdoctoral Science Foundation
(No. 2015M571126). S.K. acknowledges support from SERB-DST project
No. SR/FTP/PS-102/2011, DST FIST project No. SR/FST/MSI-090/2013(C), and the
warm hospitality and research facilities provided by the Inter-University
Center for Astronomy and Astrophysics (IUCAA), India where part of this work
was carried out. B.R. was supported in part by DOE grant DEFG 03-99EP41093.

\end{document}